%% file: main.tex
\documentclass[%
 reprint,
 letter,
 superscriptaddress,
 showpacs,preprintnumbers,
 amsmath,amssymb,
 aps,
 prl
]{revtex4-1}

\usepackage{graphicx}
\usepackage[draft]{pgf}
\usepackage[normalem]{ulem}

\newcommand{\mdot}{\ensuremath{\dot{M}_{-6}}}
\newcommand{\rhoc}{\ensuremath{\rho_{9,\mathrm{c}}}}
\newcommand{\rhocrit}{\rhoc\ensuremath{^\mathrm{crit}}}
\newcommand{\rhocritb}{\ensuremath{\rho_{\mathrm{c}}^\mathrm{crit}}}
\newcommand{\rhoign}{\rhoc\ensuremath{^\mathrm{ign}}}

\usepackage[colorlinks,citecolor=blue]{hyperref}

\begin{document}

\title{Discovery of an Exceptionally Strong $\beta$-Decay Transition of $^{20}$F\\ and Implications for the Fate of Intermediate-Mass Stars}


\author{O.~S.~Kirsebom}
\email[Corresponding author: ]{oliver.kirsebom@dal.ca}
\affiliation{Department of Physics and Astronomy, Aarhus University, DK-8000 Aarhus C, Denmark}
\affiliation{Institute for Big Data Analytics, Dalhousie University, Halifax, NS, B3H 4R2, Canada}


\author{S.~Jones} 
\affiliation{Computational Physics (XCP) Division, Los Alamos National Laboratory, New Mexico 87545, USA}
\affiliation{Heidelberger Institut f\"ur Theoretische Studien, D-69118 Heidelberg, Germany}

\author{D.~F.~Str\"omberg}
\affiliation{Institut f{\"u}r Kernphysik
  (Theoriezentrum), Technische Universit{\"a}t Darmstadt,
  Schlossgartenstra{\ss}e 2, 64289 Darmstadt, Germany}
\affiliation{GSI Helmholtzzentrum f\"ur Schwerionenforschung,
  Planckstra{\ss}e~1, 64291 Darmstadt, Germany}

\author{G.~Mart\'inez-Pinedo}
\email{g.martinez@gsi.de}
\affiliation{GSI Helmholtzzentrum f\"ur Schwerionenforschung,
  Planckstra{\ss}e~1, 64291 Darmstadt, Germany}
\affiliation{Institut f{\"u}r Kernphysik
  (Theoriezentrum), Technische Universit{\"a}t Darmstadt,
  Schlossgartenstra{\ss}e 2, 64289 Darmstadt, Germany}

\author{K.~Langanke}
\affiliation{GSI Helmholtzzentrum f\"ur Schwerionenforschung,
  Planckstra{\ss}e~1, 64291 Darmstadt, Germany}
\affiliation{Institut f{\"u}r Kernphysik
  (Theoriezentrum), Technische Universit{\"a}t Darmstadt,
  Schlossgartenstra{\ss}e 2, 64289 Darmstadt, Germany}

\author{F.~K.~R{\"o}pke}
\affiliation{Heidelberger Institut f\"ur Theoretische Studien, D-69118 Heidelberg, Germany}
\affiliation{Zentrum f\"ur Astronomie der Universit\"at Heidelberg, Institut f\"ur Theoretische Astrophysik, D-69120, Heidelberg, Germany}


\author{B.~A.~Brown}
\affiliation{National Superconducting Cyclotron Laboratory, Michigan State University, East Lansing, Michigan 48824, USA}

\author{T.~Eronen}
\affiliation{University of Jyvaskyla, Department of Physics, P.O.\ Box 35, FI-40014, University of Jyvaskyla, Finland}

\author{H.~O.~U.~Fynbo}
\affiliation{Department of Physics and Astronomy, Aarhus University, DK-8000 Aarhus C, Denmark}

\author{M.~Hukkanen}
\affiliation{University of Jyvaskyla, Department of Physics, P.O.\ Box 35, FI-40014, University of Jyvaskyla, Finland}

\author{A.~Idini}
\affiliation{Division of Mathematical Physics, Department of Physics, LTH, Lund University, P.O.\ Box 118, S-22100 Lund, Sweden}

\author{A.~Jokinen}
\affiliation{University of Jyvaskyla, Department of Physics, P.O.\ Box 35, FI-40014, University of Jyvaskyla, Finland}

\author{A.~Kankainen}
\affiliation{University of Jyvaskyla, Department of Physics, P.O.\ Box 35, FI-40014, University of Jyvaskyla, Finland}

\author{J.~Kostensalo}
\affiliation{University of Jyvaskyla, Department of Physics, P.O.\ Box 35, FI-40014, University of Jyvaskyla, Finland}

\author{I.~Moore}
\affiliation{University of Jyvaskyla, Department of Physics, P.O.\ Box 35, FI-40014, University of Jyvaskyla, Finland}

\author{H.~M\"oller}
\affiliation{GSI Helmholtzzentrum f\"ur Schwerionenforschung,
  Planckstra{\ss}e~1, 64291 Darmstadt, Germany}
\affiliation{Institut f{\"u}r Kernphysik
  (Theoriezentrum), Technische Universit{\"a}t Darmstadt,
  Schlossgartenstra{\ss}e 2, 64289 Darmstadt, Germany}

\author{S.~T.~Ohlmann}
\affiliation{Heidelberger Institut f\"ur Theoretische Studien, D-69118 Heidelberg, Germany}
\affiliation{Max Planck Computing and Data Facility, D-85748 Garching, Germany}

\author{H.~Penttil\"a}
\affiliation{University of Jyvaskyla, Department of Physics, P.O.\ Box 35, FI-40014, University of Jyvaskyla, Finland}

\author{K.~Riisager}
\affiliation{Department of Physics and Astronomy, Aarhus University, DK-8000 Aarhus C, Denmark}

\author{S.~Rinta-Antila}
\affiliation{University of Jyvaskyla, Department of Physics, P.O.\ Box 35, FI-40014, University of Jyvaskyla, Finland}

\author{P.~C.~Srivastava}
\affiliation{Department of Physics, Indian Institute of Technology, Roorkee 247667, India}

\author{J.~Suhonen}
\affiliation{University of Jyvaskyla, Department of Physics, P.O.\ Box 35, FI-40014, University of Jyvaskyla, Finland}

\author{W.~H.~Trzaska}
\affiliation{University of Jyvaskyla, Department of Physics, P.O.\ Box 35, FI-40014, University of Jyvaskyla, Finland}

\author{J.~\"Ayst\"o}
\affiliation{University of Jyvaskyla, Department of Physics, P.O.\ Box 35, FI-40014, University of Jyvaskyla, Finland}

\date{\today}

\begin{abstract}

A significant fraction of stars between 7--11 solar masses are
thought to become supernovae, but the explosion mechanism is
unclear. The answer depends critically on the rate of electron
capture on $^{20}$Ne in the degenerate oxygen-neon stellar
core. However, due to the unknown strength of the transition between
the ground states of $^{20}$Ne and $^{20}$F, it has not previously
been possible to fully constrain the rate. By measuring the
transition, we have established that its strength is exceptionally
large and enhances the capture rate by several orders of
magnitude. This has a decisive impact on the evolution of the core,
increasing the likelihood that the star is (partially) disrupted by
a thermonuclear explosion rather than collapsing to form a neutron
star.  Importantly, our measurement resolves the last remaining
nuclear physics uncertainty in the final evolution of degenerate
oxygen-neon stellar cores, allowing future studies to address the
critical role of convection, which at present is poorly understood.
\end{abstract}

\maketitle

Stars of 7--11 solar masses (M$_{\odot}$) are prevalent in the Galaxy,
their birth and death rate comparable to that of all heavier stars
combined~\cite{doherty2017}. Yet, the ultimate fate of such
``intermediate-mass stars'' remains uncertain. %
According to current
models~\cite{poelarends2008,jones2013,takahashi2013}, a significant
fraction explode, but the mechanism is a matter of ongoing
debate~\cite{isern1991,canal1992,gutierrez1996,jones2016}. %
The answer---gravitational collapse 
or thermonuclear explosion---depends critically on the rate of 
electron capture on $^{20}$Ne in the stellar core. However, 
due to the unknown strength of the transition between the ground 
states of $^{20}$Ne and $^{20}$F, it has not previously 
been possible to constrain this rate in the relevant 
temperature-density regime~\cite{pinedo2014}. Here, we report the 
first measurement of this transition, provide the first accurate 
determination of the capture rate and explore the astrophysical 
implications.

Intermediate-mass stars that undergo central carbon burning become super-AGB 
stars~\cite{doherty2017} with a degenerate oxygen-neon (ONe) core consisting 
mainly of $^{16}$O and $^{20}$Ne and smaller amounts of $^{23}$Na and 
$^{24,25}$Mg. We are interested in the scenario where the ONe core is able to 
increase its mass gradually and approach the Chandrasekhar limit, 
$M_{\mathrm{Ch}}\sim 1.37$~M$_\odot$. This can occur if nuclear burning continues 
long enough outside the core or if the core, having lost its outer 
layers, becoming a white dwarf (WD), is able to accrete material from a binary 
companion star. %
As the core approaches $M_{\mathrm{Ch}}$, it contracts and warms up, but only 
gradually as the heating from compression is balanced by cooling 
via the emission of thermal neutrinos. The density, on the other hand, rises 
rapidly eventually triggering a number of electron-capture 
processes that greatly influence the temperature evolution 
of the core. %
First, the core is cooled by cycles of electron capture followed 
by $\beta$ decay on the odd-mass nuclei $^{25}$Mg and $^{23}$Na~\cite{schwab2017}. 
At higher densities, the core is cooled by another such cycle on 
$^{25}$Na, and heated by double electron captures on the even-mass 
nuclei $^{24}$Mg and $^{20}$Ne, which produce substantial energy 
in the second capture. %
Electron capture on $^{24}$Mg occurs first at lower densities due to
its smaller $Q$-value, but $^{24}$Mg is depleted 
before the temperature can reach the threshold for oxygen ignition 
($T\sim 10^9$~K). Instead, oxygen is ignited by electron capture on 
$^{20}$Ne at somewhat higher densities. %
Previous
studies~\cite{miyaji1980,nomoto1982,nomoto1984,isern1991,canal1992,
  gutierrez1996,schwab2015,schwab2017} have considered that electron
capture on $^{20}$Ne at such conditions proceeds mainly by the allowed
transition from the ground state in $^{20}$Ne to the first $1^+$ state
in $^{20}$F, which requires a central density of the stellar core of $\rhoc \approx 9.8$
($\rho_9 \equiv \rho/{10^9}$~g~cm$^{-3}$), but it was recently
argued~\cite{pinedo2014} that electron capture on $^{20}$Ne can start
at much lower densities of $\rhoc \approx 6.8$ via the
second-forbidden, non-unique, $0^+\rightarrow 2^+$ transition connecting the ground states of $^{20}$Ne
and $^{20}$F. However, due to the transition's unknown strength it was
not possible to determine its impact~\cite{schwab2015}. %
The onset of electron capture on $^{20}$Ne heats the central region producing a
large temperature gradient, which by itself would drive convection but is 
counteracted by the composition gradient, which has a stabilizing effect. 
Stellar models are therefore sensitive to the treatment of
convection~\cite{isern1991,canal1992,hashimoto2013,tominaga2013,schwab2015}
and electron screening~\cite{gutierrez1996,schwab2015}, predicting
central oxygen ignition densities in the range $\rhoign \approx 8.9$--15.8. %

The fate of the star---gravitational collapse or thermonuclear explosion---is
sensitive to the competition between electron capture and nuclear energy
generation. If the ignition of oxygen occurs below some critical central density
$\rhocritb$, oxygen burning releases sufficient energy to reverse the collapse
and completely or partially disrupt the star in a thermonuclear
explosion~\cite{jones2016}. If it occurs above $\rhocritb$, the deleptonization
behind the burning front is so rapid that the loss in pressure cannot be
recovered by nuclear burning.  Therefore, the collapse continues to nuclear
densities, resulting in the birth of a neutron star and the ejection of the
stellar envelope~\cite{kitaura2006,janka2008}.  Stability analyses based on
spherically symmetric simulations predict $\rhocrit=8.9$~\cite{times92} though 
such one-dimensional simulations are able to produce thermonuclear explosions at 
$\rho_\mathrm{9,c}\approx 10$ if the flame propagates fast enough~\cite{NomoKondo91}. 
In fact, multi-dimensional simulations are necessary to model the flame
propagation as the efficiency of the thermonuclear combustion is set by
non-linear instabilities and turbulence that govern the flame propagation speed.
Implementing such effects in numerical schemes is very challenging. 2D
simulations predict $\rhocrit=7.9$--8.9~\cite{leung2019} while 3D
simulations still produce thermonuclear explosions at these
densities~\cite{jones2016}.  Due to the non-linear nature of the physical processes involved, the
outcome should be highly sensitive to the initial conditions. From simulations
of thermonuclear supernovae in carbon-oxygen WDs~\cite{fink2013}, we expect that
the geometry and the location of the ignition region have a significant impact
on the evolution of the flame morphology. Indeed, 2D simulations just above the
critical density no longer predict collapse if oxygen is ignited off
center~\cite{leung2019}.

This illustrates that precise knowledge of the ignition conditions 
is critical for determining the fate of these intermediate-mass stars. 
Therefore, the strength of the second-forbidden transition connecting the 
ground states of $^{20}$Ne and $^{20}$F was determined through
the measurement of the transition's branching ratio in the $\beta$
decay of $^{20}$F. %
Here we briefly summarize the main aspects of the measurement; 
details are given in an accompanying paper~\cite{kirsebom2019_prc}. %
The measurement was performed at the JYFL Accelerator Laboratory in Jyv\"askyl\"a, Finland, using a
low-energy radioactive $^{20}$F beam from the IGISOL-4 facility~\cite{arje1985, moore2013}. %
Singly-charged $^{20}$F$^+$ ions were produced by bombarding a BaF$_2$ 
target with 6-MeV deuterons. The ions were accelerated 
to 30~keV, separated according to their mass-to-charge ratio, and 
guided to the experimental station where they were implanted in 
a thin carbon foil. %
The detection system consisted of a Siegbahn-Sl{\"a}tis type 
intermediate-image magnetic electron transporter~\cite{julin1988} combined 
with a segmented plastic-scintillator detector. The magnetic transporter 
served to focus the high-energy electrons from the forbidden 
ground-state transition into the detector, while suppressing the intense 
flux of $\gamma$-rays and lower-energy electrons due to the allowed transition 
to the first-excited state in $^{20}$Ne, and hence eliminating $\beta\gamma$ 
summing and $\beta\beta$ pile-up as sources of background. Meanwhile, the 
segmentation of the detector allowed for highly efficient rejection (99.72\%) 
of the cosmic-ray background, while a baffle was used to prevent positrons 
from reaching the detector.
Finally, a LaBr$_3$(Ce) detector was used to measure the 1.63-MeV $\gamma$ 
ray associated with the allowed transition, ensuring overall normalisation
of the measurement. %

\begin{figure}%
    \centering%
    \includegraphics[width=\linewidth]{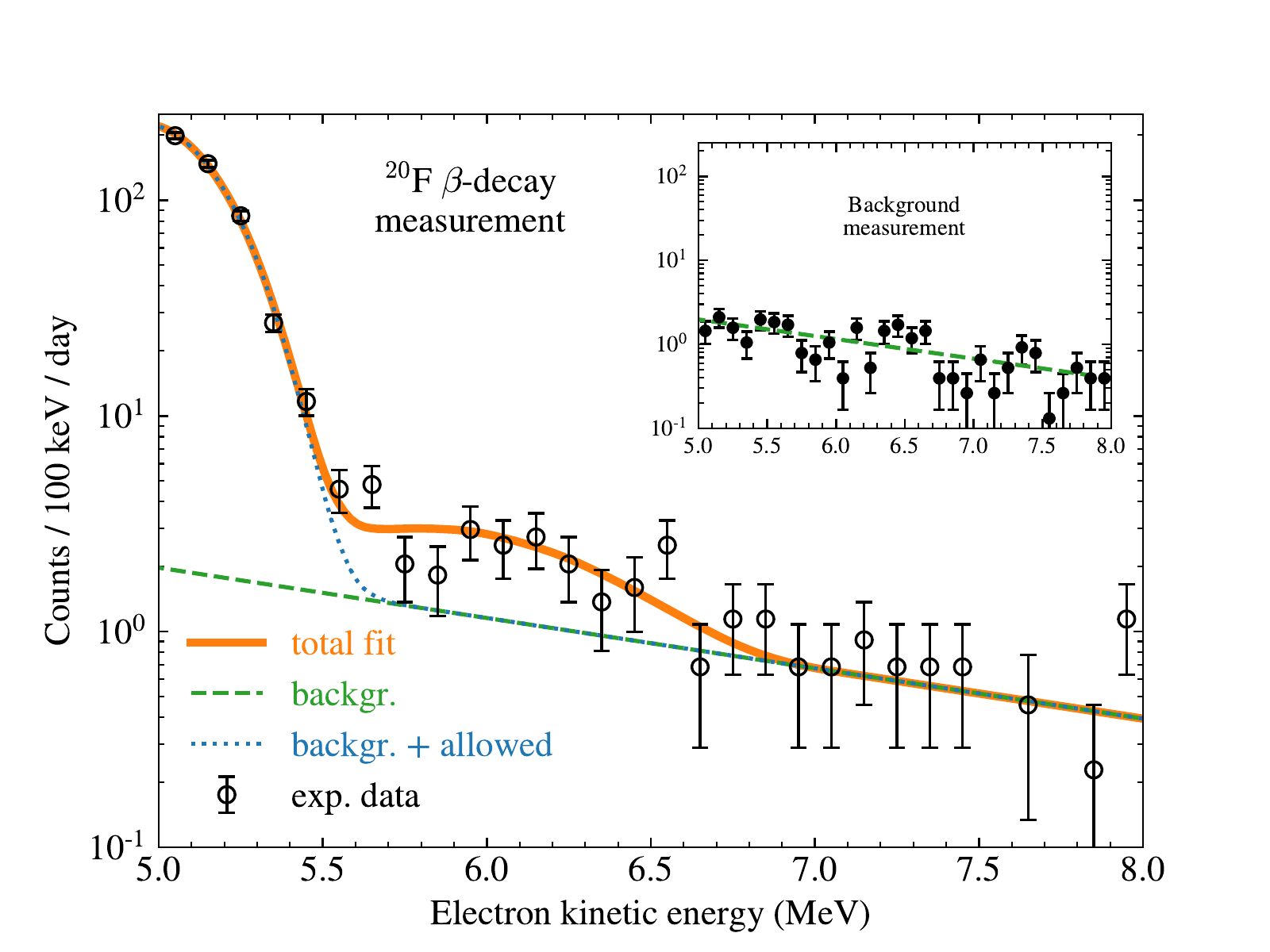}%
    \caption{\label{fig:exp} $\beta$ spectrum obtained with the
      magnetic transporter set to select the high-energy tail of the
      forbidden ground-state transition in the $\beta$ decay 
      of $^{20}$F. The inset shows the background
      spectrum obtained under the same conditions, but without the
      $^{20}$F beam. The spectrum obtained with beam exhibits a clear
      excess in the region 5.6--6.8~MeV due to the forbidden transition.}%
\end{figure}

The allowed $\beta$ spectra of $^{20}$F and $^{12}$B and monoenergetic conversion electrons from a $^{207}$Bi 
source were used to characterize the acceptance window of the magnetic transporter and 
the response of the plastic-scintillator detector for electron energies up to 8.0~MeV.
This permitted the detection efficiency of the forbidden transition to be determined 
directly from experimental data with a precision of 16\%. %
The response was further modelled with a GEANT4 simulation~\cite{geant2003,geant2016} and good 
agreement was achieved between measured and simulated energy distributions. %
For the measurement of the forbidden transition, data were collected for 105 hours 
with the magnet tuned to focus electrons with energies of $\sim 6$--7~MeV 
and background data were collected for 183 hours without beam, but with the magnet still on. The $\beta$ spectra obtained in these long 
measurements are displayed in Fig.~\ref{fig:exp}. %
The forbidden transition (end-point energy of 7.025~MeV) gives rise to the excess
counts between 5.6--6.8~MeV, while the five orders of magnitude
more intense allowed transition to the
first-excited state in $^{20}$Ne (end-point energy of 5.391~MeV)
dominates at lower energies. 

The statistically significant detection of a signal was established
through a maximum likelihood fit in which the shapes of the allowed
and forbidden transitions were obtained from the GEANT4 simulation,
while the shape of the cosmic-ray background was parameterized by an
exponential function with two free parameters. %
Including the forbidden transition in the fit model, we obtained a satisfactory 
fit quality ($p$-value of 0.080) and constrained the magnitude of 
the signal with a statistical uncertainty of 19\%. In contrast, fitting without 
the forbidden transition gives an unsatisfactory fit quality ($p$-value of 0.0003). %
Correcting for the $\beta$ detection efficiency, normalizing to the
total number of decays inferred from the 1.63~MeV $\gamma$-ray
yield, and adopting the shape factor predicted by our shell-model
calculation (see below), we determine the branching ratio to be
$0.41(11)\times 10^{-5}$, where systematical and statistical
uncertainties have been added in quadrature. %
Using the known half-life for $^{20}$F of 11.0062(80)~s~\cite{burdette2019}, we 
determine the transition strength to be $\log ft = 10.89(11)$.
Thus, the transition is three orders of magnitude stronger than the
only other known second-forbidden, non-unique transition for a nucleus
with a similar mass ($^{36}\mathrm{Cl}\rightarrow{}^{36}\mathrm{Ar}$, 
$\log ft = 13.321(3)$~\cite{kriss2004}) and, in fact, one of the strongest 
of its kind~\cite{singh1998}. %

\begin{figure}%
    \centering
    \includegraphics[width=\linewidth]{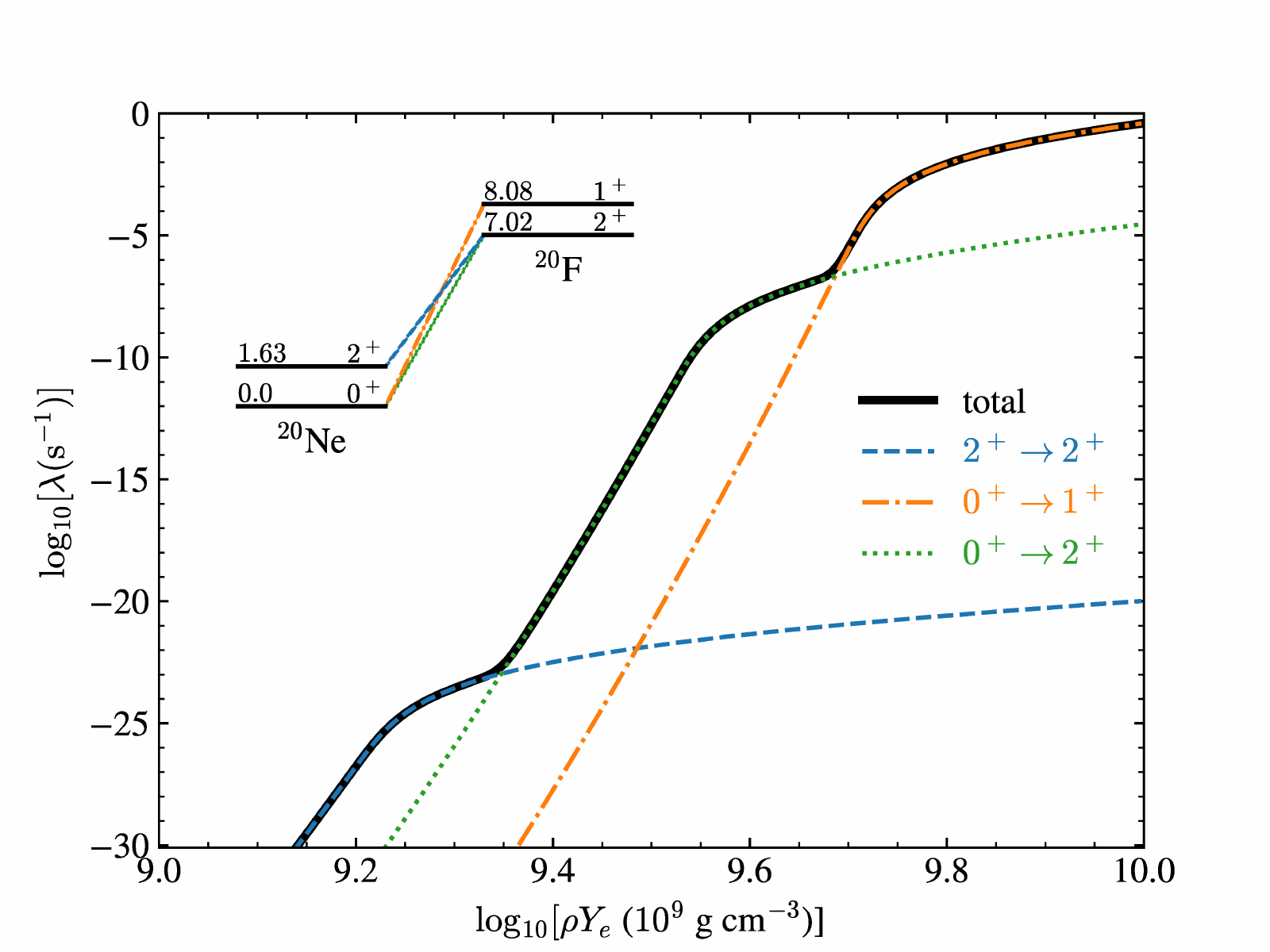}
    \caption{\label{fig:rate}Astrophysical electron-capture rate as a function of density for a temperature of $T=0.4$~GK.
            A simplified level scheme shows the main transitions with the nuclear levels labeled by their spin-parity and 
            energy in MeV relative to the $^{20}$Ne ground state.}
\end{figure}

The electron-capture rate on $^{20}$Ne is shown in Fig.~\ref{fig:rate} for a 
temperature of $T=0.4$~GK.
%
%
Including the forbidden transition, the electron capture
rate increases by up to eight orders of magnitude in the important
density range $\rho_9 \simeq 4.5$--9.5 ($\log_{10}[\rho Y_e(\mathrm{g}\;\mathrm{cm}^{-3})] \simeq 9.35$--9.68). %
As a result, it competes with the timescale 
of core contraction and affects the evolution of the core. %
We note that if the strength of the forbidden transition had been 
similar to what is observed for $^{36}$Cl, the electron-capture rate would ``only'' 
have been enhanced by five orders of magnitude. It would then have remained below the contraction rate, and 
the forbidden transition would not have been able to affect the evolution 
of the stellar core.

The electron-capture rate and $\beta$-decay rates were calculated
following the approach of Ref.~\cite{pinedo2014}. For forbidden
transitions, the constant matrix element is replaced by an energy
dependent shape factor~\cite{Behrens.Buehring:1982} that is a function
of the matrix elements between the initial and final nuclear
states. The exact relationship depends on the type of transition. We
use the formalism of
Refs.~\cite{Behrens.Buehring:1971,Behrens.Buehring:1982} for $\beta^-$
and electron capture. The matrix elements are determined from
shell-model calculations performed in the $sd$ shell with the USDB
interaction~\cite{brown2006} using harmonic oscillator single-particle
wave functions and constrained by the known strength of the analog
E2 transition in $^{20}$Ne together with the conserved vector
current theory. Moreover, we use the bare value of the axial coupling
constant since previous calculations of unique second-forbidden
transitions have not found evidence of quenching of the axial coupling
constant~\cite{Warburton:1992,Martinez-Pinedo.Vogel:1998}.  Our
calculations reproduce the observed half-life of the second-forbidden
transition to within better than 10\%.  The matrix elements, rescaled
to the observed half-life, are then used for the evaluation of the
electron-capture rate taking into account the appropriate
kinematics. In this way, we are able to constrain the electron-capture
rate to within 25\% at the relevant density and temperature conditions
taking into account also the uncertainty on the theoretical shape
factor~\cite{kirsebom2019_prc}.

\begin{figure}
  \includegraphics[width=\linewidth]{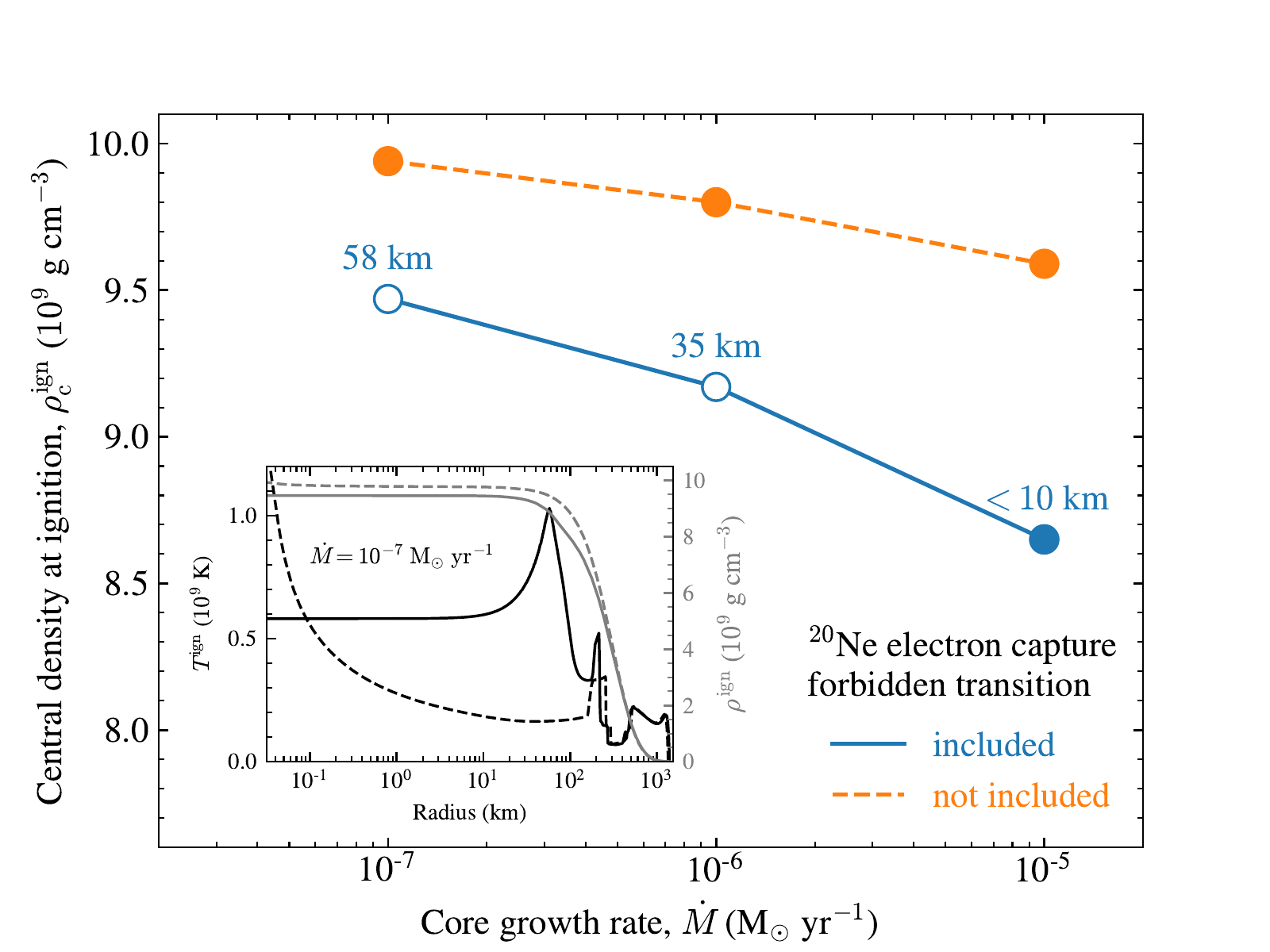}
  \caption{\label{fig:mesa} Central ignition density 
  vs.\ growth rate for a contracting, degenerate ONe core, with 
  and without the forbidden transition between the ground states 
  of $^{20}$Ne and $^{20}$F. Filled circles denote cases in which 
  oxygen ignition occurs centrally, while empty circles denote 
  off-center ignition at the indicated radius. The panel shows 
  temperature and density profiles at the time of ignition for 
  the low growth rate ($10^{-7}\;$M$_\odot\;$yr$^{-1}$).}
\end{figure}

To quantify the impact of the forbidden transition, we simulate the final evolution of an accreting
ONe core using the stellar evolution code MESA~\cite{paxton2018}
following the procedure of Refs.~\cite{schwab2015,schwab2017} where matter is accreted onto the core
at a constant rate, $\dot{M}$. We consider the cases
$\dot{M}_{-6} = 0.1$, 1.0 and 10
($\mdot \equiv \dot{M} / 10^{-6}\;$M$_\odot\;$yr$^{-1}$)
representative of thermally stable hydrogen burning
$(\mdot\approx 0.4\textrm{--}0.7)$~\cite{wolf2013} and helium burning
$(\mdot\approx 1.5\textrm{--}4.5)$~\cite{brooks2016}. %
We find that the inclusion of the forbidden transition allows the 
electron captures on $^{20}$Ne to proceed at lower densities (see Supplemental Material). 
However, since the forbidden transition is more than five
orders of magnitude weaker than a typical allowed transition, the
captures do not produce a thermal runaway, as would be the case for an
allowed transition, but rather a gradual heating of the core. %
As a result, the star develops an isothermal core with a radius of 10--60~km 
and for the $\mdot=0.1$ and $1.0$ cases, this phase lasts long enough that 
most $^{20}$Ne within the isothermal core is converted to $^{20}$O by
double electron capture. 
Hence, further heating occurs in the outer regions of the core triggering an
off-center ignition of oxygen. For the $\mdot=10$ case, the ignition
occurs in a central region with $10$~km radius. 
Fig.~\ref{fig:mesa} summarizes the results of our simulations.  For
all cases considered, the contribution of the forbidden transition
leads to earlier heating resulting in oxygen ignition at lower
densities. Changes in the
chemical composition, in particular the initial amount of $^{24}$Mg
and $^{25}$Mg, affect the evolution somewhat, but do not alter the
picture qualitatively, unless the $^{24}$Mg fraction is made very
large~\cite{schwab2015}.

\begin{figure}
  \includegraphics[width=\linewidth]{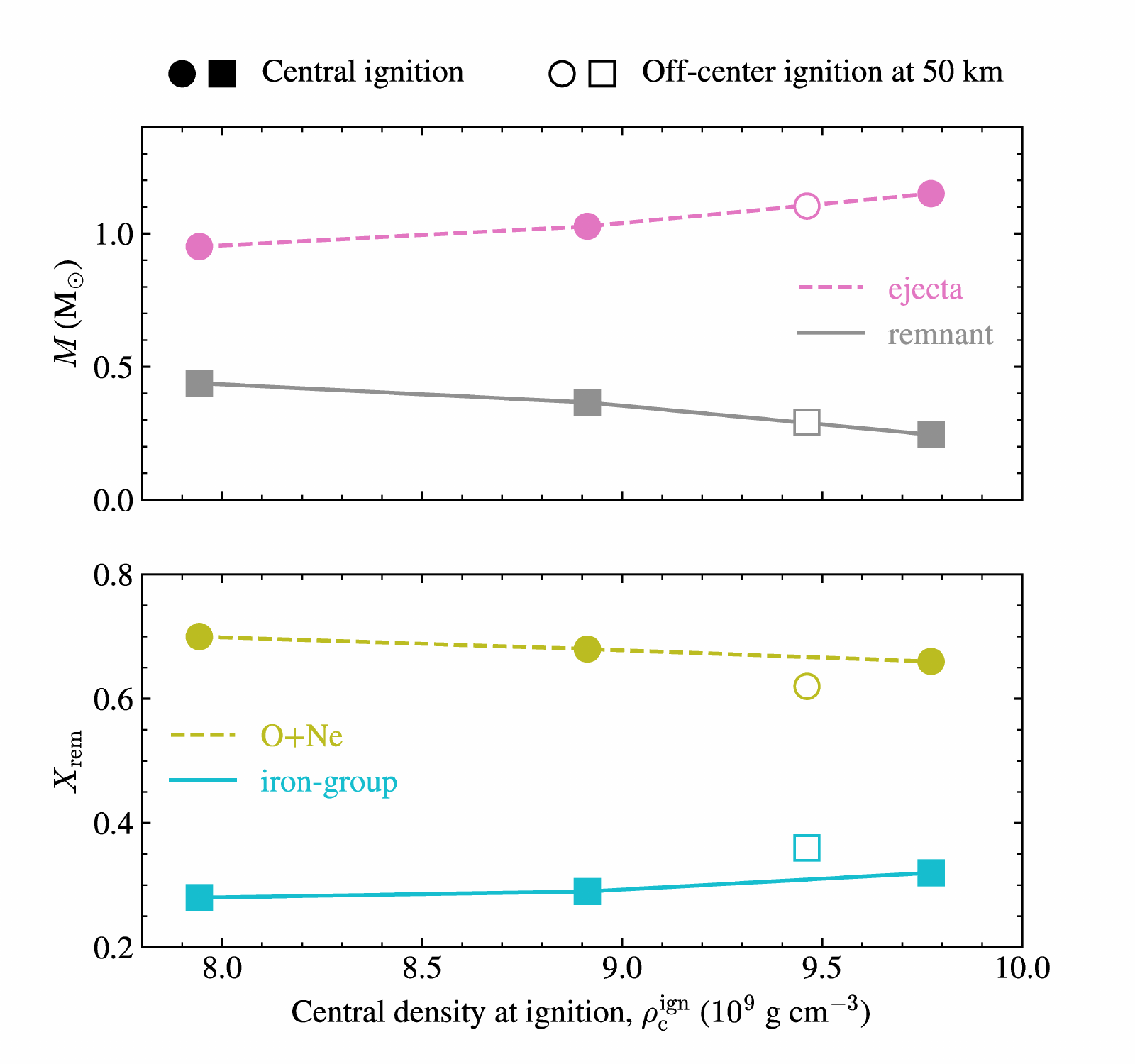}
  \caption{\label{fig:hydro_panels} Mass ($M$) of bound remnant and ejecta and 
    mass fractions ($X$) of oxygen + neon and iron-group elements 
    in the remnant are shown as a function of the central
    density at ignition ($\rho_{\mathrm{c}}^{\mathrm{ign}}$). Filled markers denote
    simulations with central ignition; empty markers denote simulations 
    with ignition occurring in a sphere with radius of 50~km.}
\end{figure}

Determining the final outcome after oxygen ignition---gravitational
collapse or thermonuclear explosion---requires multi-dimensional 
hydrodynamical simulations. We have performed four high-resolution 
3D hydrodynamical simulations using the LEAFS 
code~\cite{reinecke2002,jones2016} with different
assumptions for the initial density and flame geometry 
motivated by the results of the MESA stellar evolution 
simulations. We also calculate the nucleosynthesis in the 
ejecta following the approach of Ref.~\cite{jones2018}. %
None of our simulations actually result in core collapse into a
neutron star; all are partial thermonuclear explosions that produce a
bound remnant consisting of oxygen, neon and iron-group elements
(ONeFe WD). %
The inclusion of the forbidden transition, which favors an 
off-center ignition at lower densities, has a significant 
impact on the explosion: The lower density slows down the 
conductive flame and leads to less energetic burning, which 
results in a more massive remnant because less material 
is ejected (Fig.~\ref{fig:hydro_panels}, top panel). 
On the other hand, the off-center ignition leads to more 
energetic burning during the first 1~second of the 
explosion (see Supplemental Material), resulting in a higher fraction of 
iron-group elements in the remnant compared to the centrally 
ignited models (Fig.~\ref{fig:hydro_panels}, bottom panel). %

\begin{figure}
\centering
    \includegraphics[width=\linewidth]{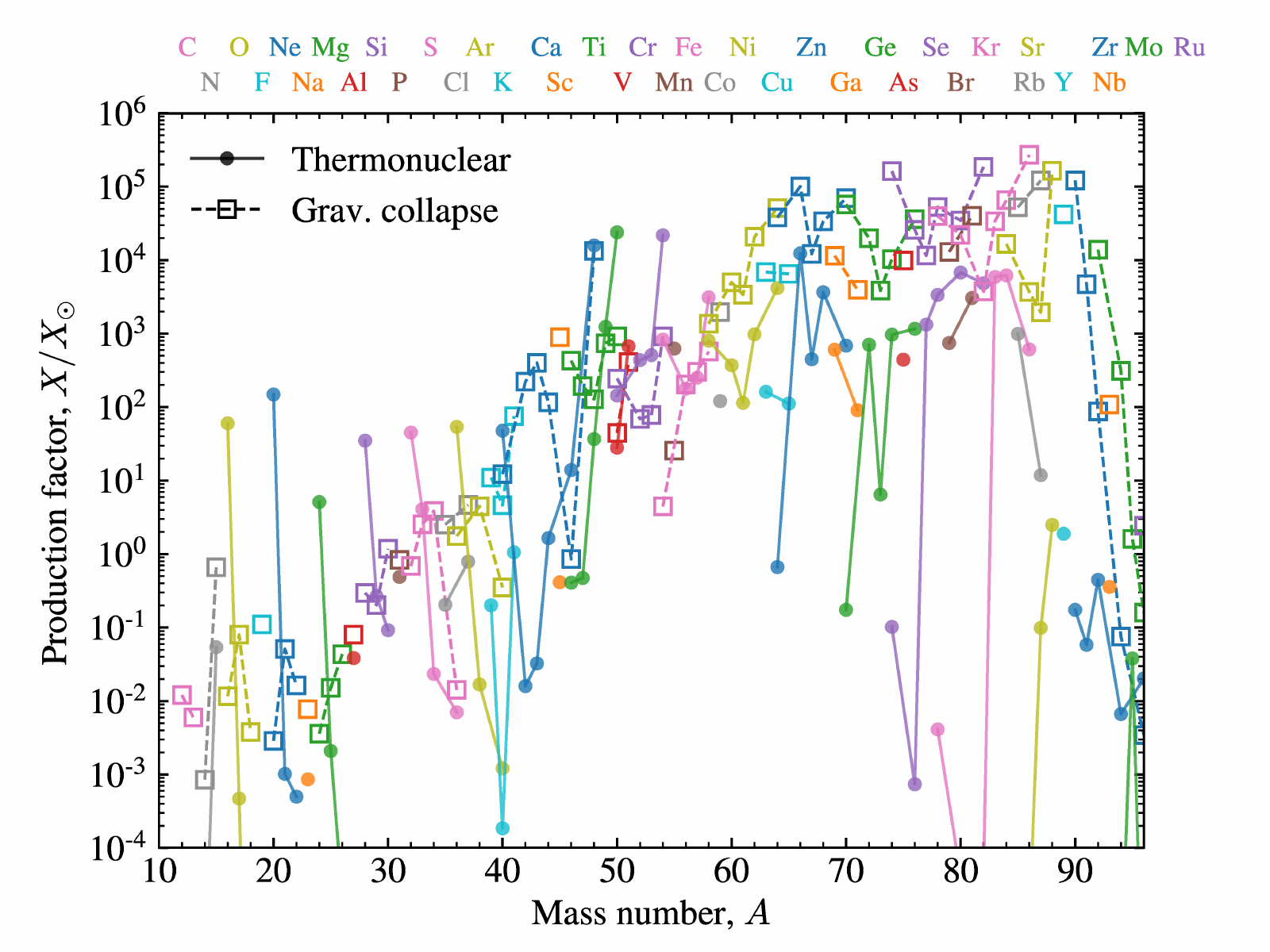}
    \caption{\label{fig:hydro} Mass fraction relative to solar, $X/X_{\odot}$, of 
    stable isotopes in the ejecta of the (off-center)     
    thermonuclear explosion compared to the gravitational 
    collapse of Ref.~\cite{wanajo2018}.}
\end{figure}

We find that the explosion mechanism has a significant impact on the 
nucleosynthesis yields. This is primarily due to thermonuclear 
explosion ejecting far more material, $M_{\mathrm{ej}}\sim 1\,$M$_{\odot}$, than the gravitational collapse, $M_{\mathrm{ej}}\sim 0.01\,$M$_{\odot}$~\cite{wanajo2018}, although the isotopic distributions also exhibit some differences (Fig.~\ref{fig:hydro}), notably in the production factors 
of $^{50}$Ti and $^{54}$Cr, which are enhanced by factors of $\sim 20$ 
in the thermonuclear explosion.
On the other hand, the changes in ignition density and geometry 
caused by the forbidden transition have a modest impact on nucleosynthesis, leading to changes of up to $\sim 10\%$ in 
the production factors of individual isotopes (see Supplemental 
Material). %
We find that the ejecta of the thermonuclear explosion are 
particularly rich in the neutron-rich isotopes $^{48}$Ca, 
$^{50}$Ti and $^{54}$Cr and the trans-iron elements 
Zn, Se and Kr (Fig.~\ref{fig:hydro}). %
This has important implications for our understanding
of early Galactic chemical evolution~\cite{jones2018} and 
may also explain unusual Ti
and Cr isotopic ratios found in presolar grains~\cite{jones2018,nittler2018}.
The radionuclide $^{60}$Fe is also produced in large amounts
($3.63\times 10^{-3}$~M$_{\odot}$), implying
that the live $^{60}$Fe found in deep-sea sediments~\cite{wallner2016}
could have originated from the recent death of a nearby
intermediate-mass star~\cite{wanajo2013fe60}. On the other hand, 
the production of $^{26}$Al is rather modest, resulting in a large 
$^{60}$Fe$/^{26}$Al ratio~\cite{jones2018}.

In summary, our work indicates that the ONe core, for realistic growth rates and 
composition, will not collapse to a neutron star, but rather be partially 
disrupted by the oxygen deflagration wave, producing a ONeFe WD and a subluminous 
Type Ia supernova\@. 
This is contrary to the commonly accepted view that 
collapse to a neutron star is more likely~\cite{gutierrez1996, leung2019} 
and has the notable corollary that the Crab Nebula (SN 1054) 
likely was the result of a low-mass iron core-collapse supernova. 
Our findings suggest that intermediate-mass stars may be an important (and 
potentially the only) channel for making ONeFe WDs. Detection or non-detection 
of such objects with future missions would provide important 
insights into the explosion mechanism. %

The present determination of the electron capture rate on $^{20}$Ne removes 
the last remaining nuclear physics uncertainty in the evolution of degenerate ONe 
cores. Not only does the new accurate capture rate result in a reduced 
ignition density below $\rhocritb$, it also modifies the 
initial conditions by causing an off center ignition. %
With this result, the most uncertain aspect of the progenitor evolution 
is whether or not the core becomes convectively unstable~\cite{schwab2017}, 
and whether the convective energy transport is efficient enough to delay 
the ignition and the start of the oxygen deflagration wave to densities 
above the critical density for collapse. Future efforts should therefore 
focus on characterising convection in the progenitor evolution. %
However, the main result of this work will not change: The new accurate 
$^{20}$Ne capture rate tips the balance in favour of a thermonuclear explosion. %

This is the first astrophysical case in which a second-forbidden 
transition has been found to play a decisive role. %
Our result allows advances in our understanding of the fate of intermediate-mass 
stars and their contribution to galactic chemical evolution, populations of 
compact objects in the Universe, and diversity of supernova light curves.

\begin{acknowledgments}

We are indebted to the technical staff at the JYFL laboratory 
and Aarhus University for their assistance with refurbishing 
the spectrometer and to the members of the IGISOL-4 group for 
their support during the experiment. %
This work has been supported by the Academy of Finland under the 
Finnish Centre of Excellence Programme 2012--2017 (Nuclear and 
Accelerator Based Physics Research at JYFL) and the Academy of 
Finland grants No.\ 275389, 284516 and 312544. %
This work was supported by the US Department of Energy LDRD program
through the Los Alamos National Laboratory. Los Alamos National
Laboratory is operated by Triad National Security, LLC, for the National
Nuclear Security Administration of U.S. Department of Energy (Contract
No. 89233218NCA000001). SJ acknowledges support from a Director's
Fellowship at Los Alamos National Laboratory.
The work of FR, SJ and STO was supported by the Klaus Tschira 
Foundation and FR received additional support through the Collaborative 
Research Center SFB 881 ``The Milky Way System'' of the German 
Research Foundation (DFG). %
BAB acknowledges the support of NSF grant PHY-1811855, and 
OSK acknowledges support from the Villum Foundation through 
Project No.\ 10117. %
DFS and GMP acknowledge the support of the Deutsche
Forschungsgemeinschaft (DFG, German Research Foundation) -
Projektnummer 279384907 - SFB 1245 ``Nuclei: From Fundamental
Interactions to Structure and Stars''; and the ChETEC COST action
(CA16117), funded by COST (European Cooperation in Science and
Technology). %
JK acknowledges the financial support of the Jenny and Antti Wihuri Foundation.

\end{acknowledgments}



%

\end{document}


\begin{center}
Supplemental material for

\vspace{0.5cm}

{\bf%
Discovery of an Exceptionally Strong $\beta$-Decay Transition of $^{20}$F\\ 
and Implications for the Fate of Intermediate-Mass Stars}

\vspace{0.5cm}

{\small%
O.~S.~Kirsebom, 
S.~Jones, 
D.~F.~Str\"omberg,
G.~Mart\'inez-Pinedo,
K.~Langanke,\\
F.~K.~R{\"o}pke, 
B.~A.~Brown,
T.~Eronen,
H.~O.~U.~Fynbo,
M.~Hukkanen,\\
A.~Idini,
A.~Jokinen,
A.~Kankainen, 
J.~Kostensalo,
I.~Moore,
H.~M\"oller,
S.~T.~Ohlmann,
H.~Penttil\"a, 
K.~Riisager,
S.~Rinta-Antila,\\
P.~C.~Srivastava,
J.~Suhonen,
W.~H.~Trzaska,
J.~\"Ayst\"o }

\vspace{0.5cm}

\end{center}

\section*{MESA stellar evolution simulations}

\begin{figure}[p]
    \centering
    \includegraphics[width=\linewidth]{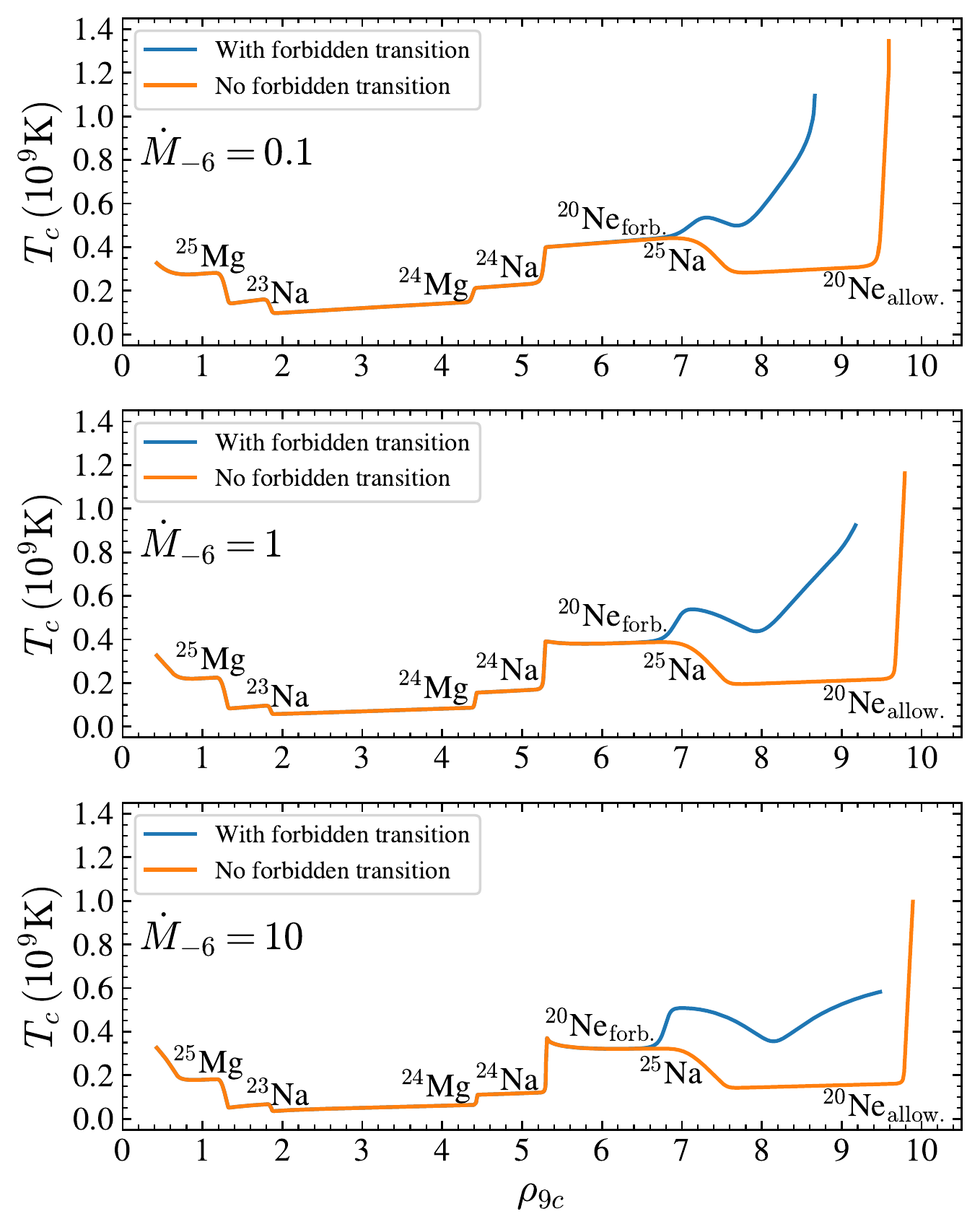}
    \caption{\label{fig:S4} 
            Central temperature as a function of central density for the MESA simulations performed 
            with the SQB15+ composition and mass-growth rates of $\mdot=0.1,1,10$ ($\mdot \equiv \dot{M} / 10^{-6}\;$M$_\odot\;$yr$^{-1}$). The onset of each electron capture reaction is labeled with the parent nucleus.}
\end{figure}

The MESA simulations follow the 
approach of Refs.~\cite{schwab2017,schwab2015}: 
The ONe core is prepared at a central density of $\rhoc = 0.4$, 
corresponding to a mass of 1.3M$_\odot$, with the SQB15+ 
composition of Ref.~\cite{schwab2017}: 
$X(^{16}\text{O})=0.50$, $X(^{20}\text{Ne})=0.390$, 
$X(^{23}\text{Na})=0.05$, $X(^{24}\text{Mg})=0.05$ 
and $X(^{25}\text{Mg})=0.01$. %
The evolution of the central density and temperature for the 
three core growth rates considered ($\dot{M}_{-6}=0.1,1,10$) 
are shown in Fig.~\ref{fig:S4}. %
Immediately following the start of the simulation, at $\rhoc = 0.4$, 
the core cools down until the compressional heating balances the 
thermal neutrino losses. The slower the core growth, the lower the 
temperature where this happens. %
When the density reaches $\rhoc = 1.3$ and 1.9, the URCA cycles 
$^{25}$Mg $\leftrightarrow$ $^{25}$Na and $^{23}$Na $\leftrightarrow$ 
$^{23}$Ne are able to operate, cooling the core to below $10^8$~K. 
An URCA cycle consists of an electron capture followed by the reverse 
$\beta$ decay; it produces two neutrinos, which escape from the core, 
carrying away significant amounts of energy. 
The core, now cooled to below $10^8$~K, continues to contract, but 
does so adiabatically because the production of thermal neutrinos 
is insignificant below $10^8$~K. %
This phase lasts until the exothermic electron captures on $^{24}$Mg 
and $^{24}$Na at $\rhoc = 4.4$ and 5.3 bring the core back to a 
trajectory set by the balance between compressional heating and 
thermal neutrino losses. %
At this point all our models have very similar central temperatures,
$T_c\approx 0.38$~GK, and a compression timescale of 
$t_{\text{compress}} \approx 8\times 10^3/ \mdot$~yr.

Models that neglect the contribution of the forbidden transition
further cool down by the URCA cycle $^{25}$Na $\leftrightarrow$
$^{25}$Ne when the core reaches $\rhoc = 7.0$, which
enhances the temperature differences between models. 
The core now evolves adiabatically~\cite{schwab2017} until the density
becomes high enough for electron captures on $^{20}$Ne to set in, 
producing a thermal runaway in the center of the star that is followed 
by the ignition of oxygen burning. %
Due to the strong sensitivity of subthreshold electron capture rates 
to temperature~\cite{pinedo2014}, the onset occurs at lower densities 
for models with higher temperatures. 

\begin{figure}[p]
    \centering
    \includegraphics[width=0.49\linewidth]{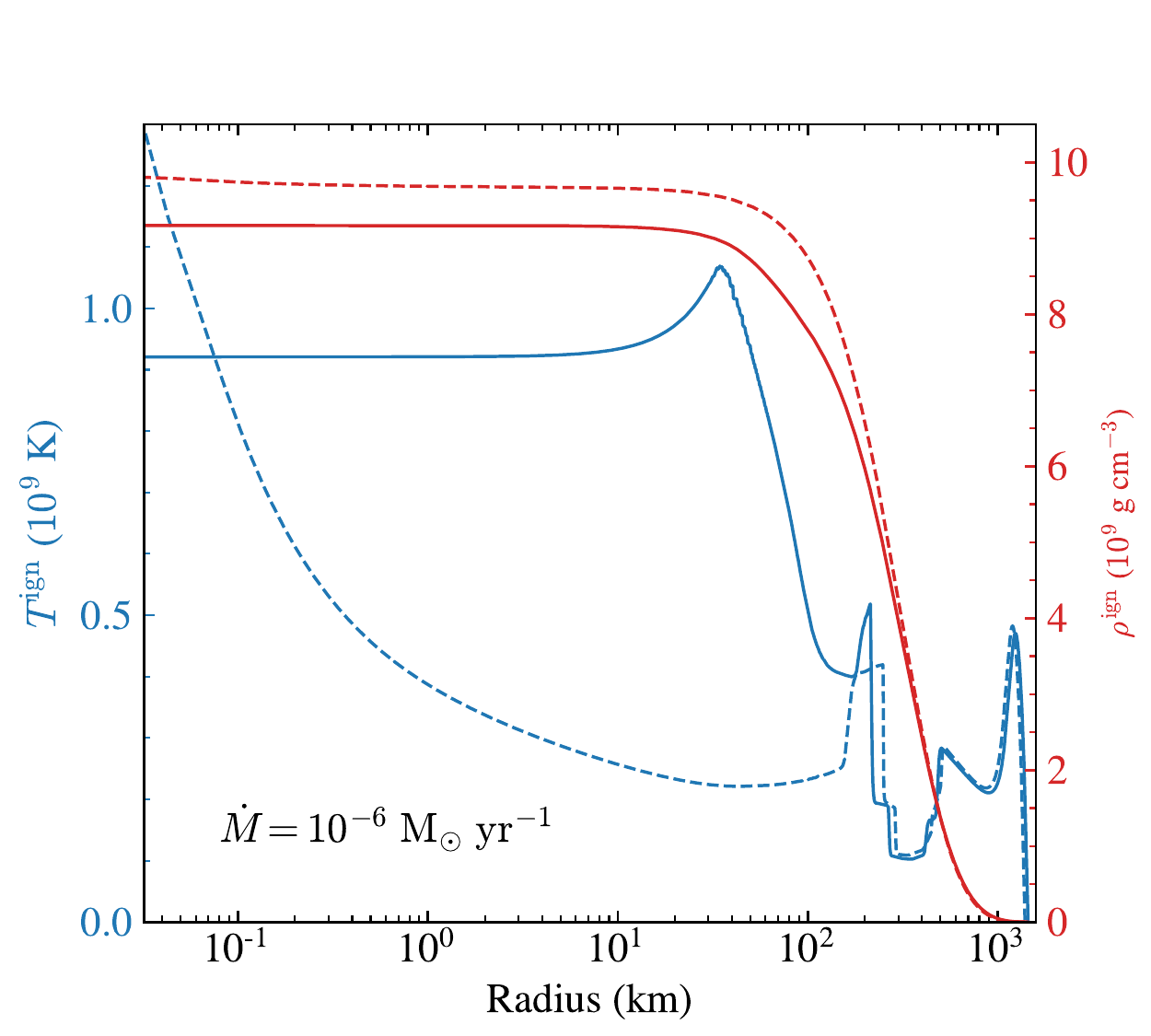}%
    \hspace{0.01\linewidth}%
    \includegraphics[width=0.49\linewidth]{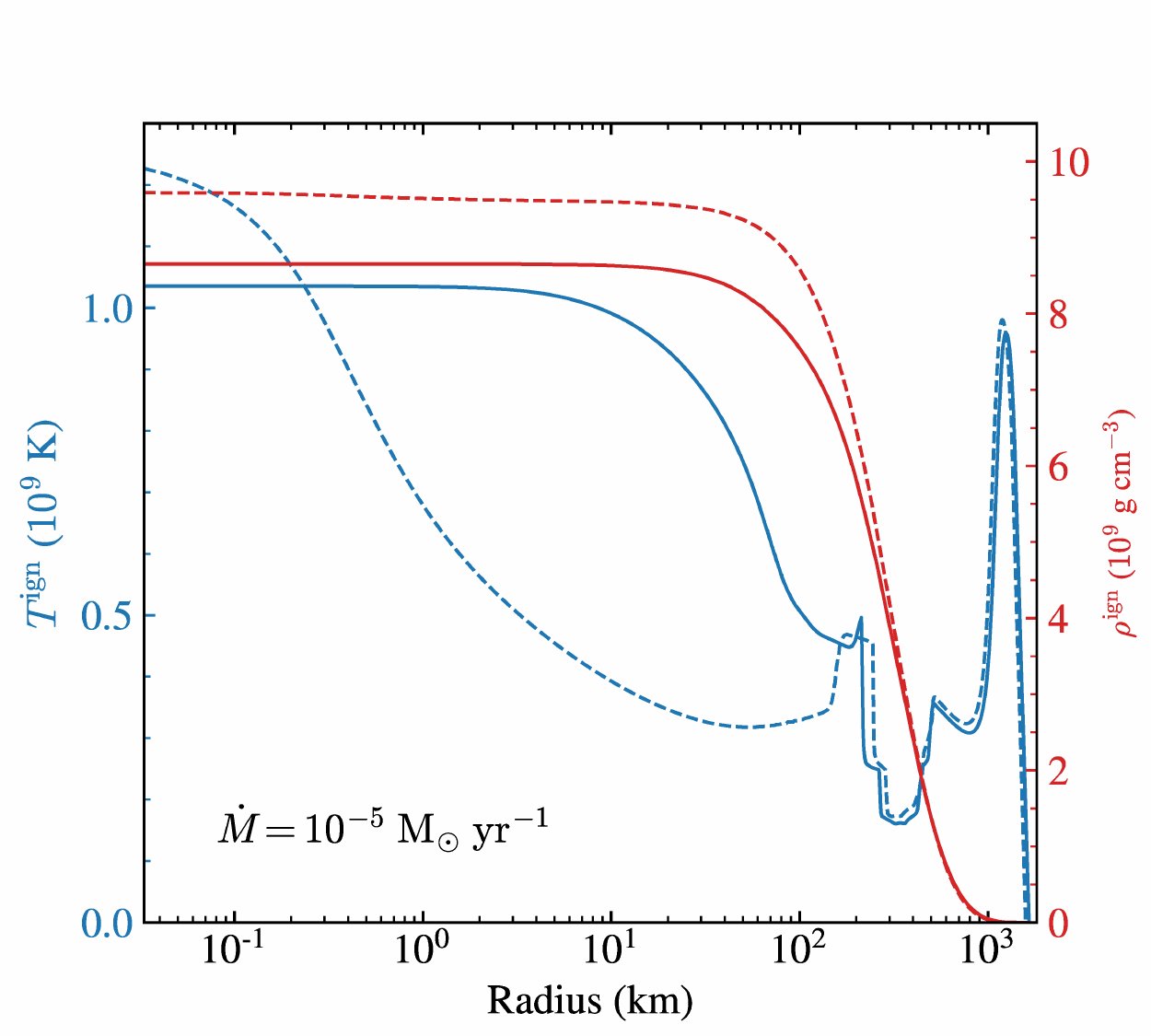}
    \caption{\label{fig:S5}Temperature and density profiles of the
      ONe core at the moment oxygen is ignited for the mass-growth
      rates $\mdot=1$ and $10$. The solid and dashed lines show the profiles 
      obtained with and without the forbidden transition, respectively.}
\end{figure}

The evolution is rather different for models that include the
forbidden transition: The $Q$ value for the $^{20}\text{Ne}\rightarrow{}^{20}$F forbidden
transition, 7.535~MeV, is slightly lower than for the URCA
pair \mbox{$^{25}$Na $\leftrightarrow$ $^{25}$Ne}, 7.744~MeV, and hence the forbidden transition can
proceed at slightly lower densities, resulting in a gradual heating 
of the core.  %
Once the URCA cycle $^{25}$Na $\leftrightarrow$ $^{25}$Ne starts to 
operate, its cooling dominates over the heating from the $^{20}$Ne 
electron captures, but when $^{25}$Na has been exhausted in the centre, 
the heating resumes. The forbidden transition continues to operate 
until the temperature and density 
reach a point where the allowed transition sets in, producing 
a thermal runaway that ignites oxygen burning. %
Note that since the ignition happens off-centre for $\mdot=1$ and 0.1,
the central temperature plotted in Fig.~\ref{fig:S4} is much
lower than the temperature at the ignition point for these two
cases. This is particularly obvious in the latter case where the
central temperature does not exceed $\sim 0.6$~GK despite oxygen
burning taking place $\sim 50$~km from the centre. %
In Fig.~\ref{fig:S5} we show the temperature and density profiles 
of the ONe core at the moment of ignition for $\mdot=1$ and 10. The profiles 
for $\mdot=0.1$ are shown in Fig.~3 of the main text.

\section*{LEAFS explosion simulations}

The 3D hydrodynamic simulations performed for this work use the
LEAFS~\cite{reinecke2002} code, updated in Ref.~\cite{jones2016} for simulating
ONe deflagrations (as opposed to CO deflagrations). LEAFS uses a uniform
Cartesian mesh for the flame and its ashes nested inside of a non-uniform
Cartesian mesh for the external structure of the star. Both meshes are allowed
to expand, exchanging cells between the two grids where necessary.  The
simulations begin from isothermal ONe white dwarfs with composition
$X(^{16}\text{O})=0.65$, $X(^{20}\text{Ne})=0.35$ and electron fraction
$Y_\text{e}=0.493$, integrated into hydrostatic equilibrium on a spherically
symmetric grid from a given central density and then mapped onto the 3D
Cartesian mesh. %

We note that the composition assumed here is slightly different from the 
composition of our 1D MESA simulations. We expect these small changes
to have a minor effect on the critical ignition density and on the resulting 
nucleosynthesis, but this requires further detailed simulations to quantify 
precisely.  In general, the composition depends on the mass of the progenitor 
star at birth and the nuclear reaction cross sections employed. %
We also note that the simulations do not include a hydrogen or helium envelope, which
would be present in a super-AGB star or stripped-envelope massive star,
respectively. This omission is unimportant for the explosion dynamics but would
substantially affect some of the observed properties of the supernova. %

The initial flame geometry for the ``centrally'' ignited simulations, as in
Ref.~\cite{jones2016}, was 300 spherical bubbles non-uniformly distributed
within a sphere of radius 50~km. The initial flame geometry for the simulations
with an off-centre ignition consisted of 96 bubbles with radius 4~km. The
bubbles were distributed evenly in polar and azymutal, $(\theta$, $\phi)$,
coordinates with 8 points in $\theta$ and 12 points in $\phi$. Each bubble's radial position
was perturbed randomly from 50~km by up to 10\%.  The procedure for
following nuclear burning, burning front progression, laminar and turbulent
flame speed calculations and deleptonization is described in more detail in
Ref.~\cite{jones2016} and references therein. Here we include a table
(Table 1) with all the relevant values obtained from the 3D
simulations. %
Here, C$_{9.77}$ corresponds to the MESA models without the forbidden transition, while
O$_{9.46}$ and C$_{8.91}$ correspond to the MESA models where the forbidden transition is
included and the core growth rate is low/intermediate and high, respectively.
For comparison, we also include model C$_{7.94}$ from Ref.~\cite{jones2016}. %

\begin{figure}[p]
\centering
    \includegraphics[width=\linewidth]{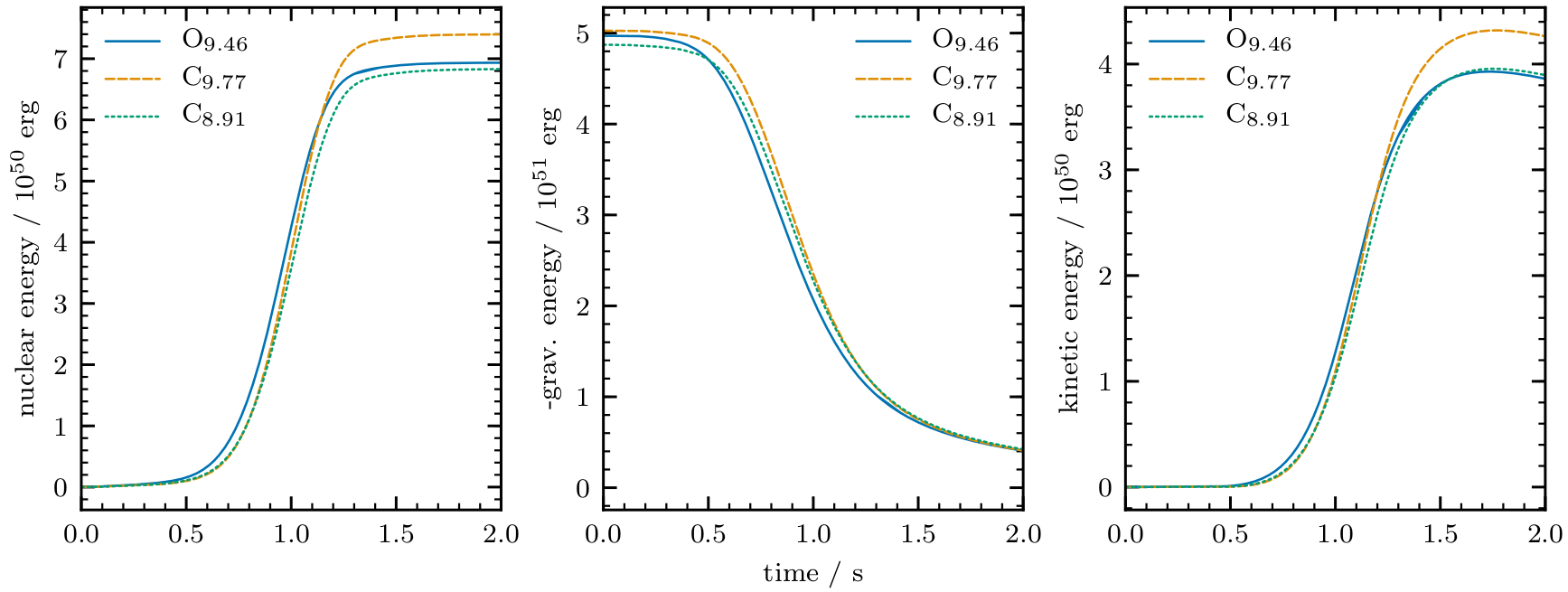}
    \caption{\label{fig:S6}
      Energy evolution of the 3D hydrodynamical
      simulations. Released nuclear energy (left),
      gravitational energy (center) and kinetic energy (right)
      as a function of the explosion time in the simulations C$_{8.91}$ 
      (short dash), O$_{9.46}$ (solid) and C$_{9.77}$ (long dash).}
\end{figure}

\begin{figure}[p]
\centering
    \includegraphics[width=\linewidth]{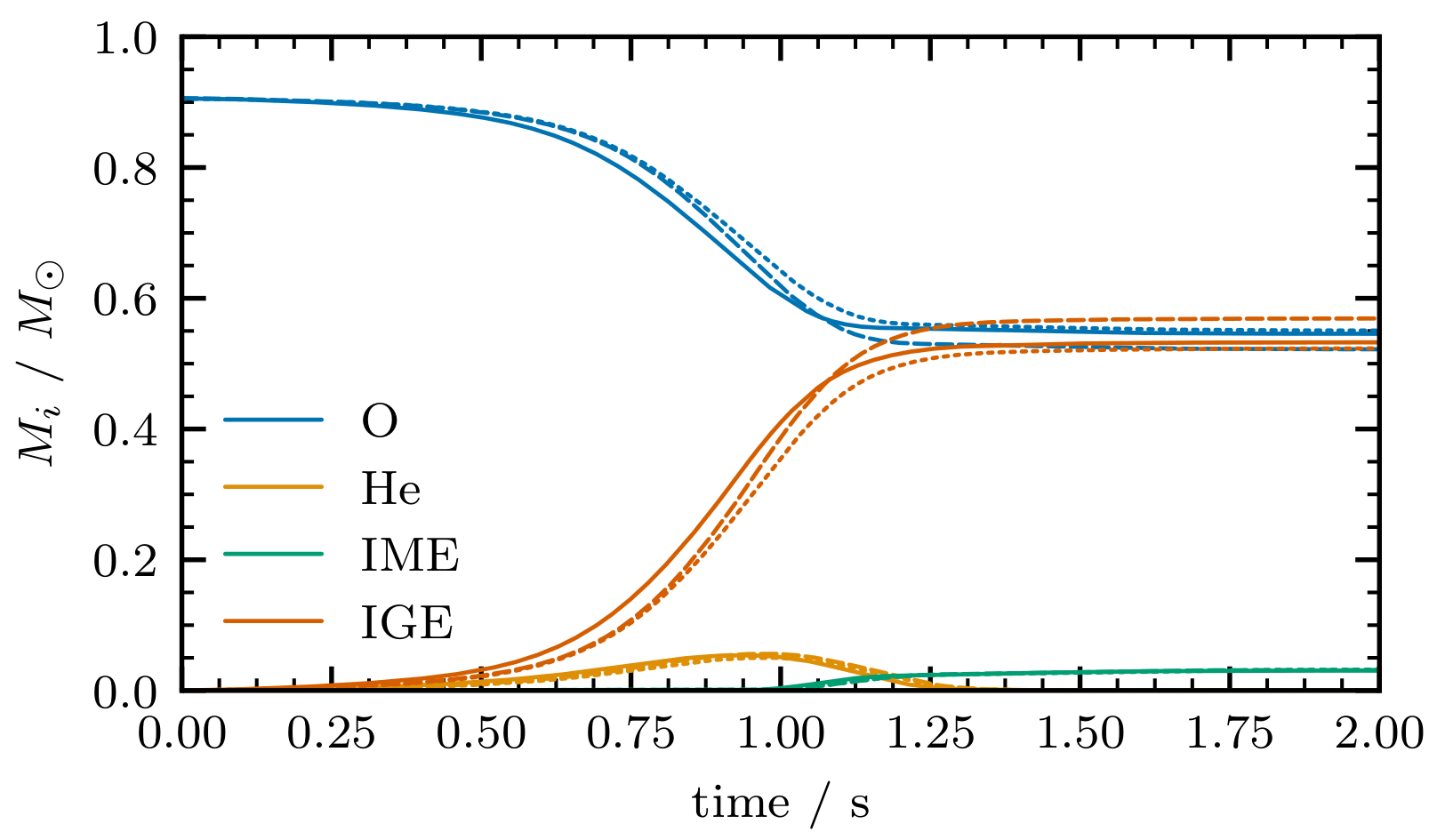}
    \caption{\label{fig:S7}Chemical evolution of the 3D hydrodynamical
      simulations. Chemical composition as a function
      of the explosion time in the simulations C$_{8.91}$ 
      (short dash), O$_{9.46}$ (solid) and C$_{9.77}$ (long dash). 
      IME: Intermediate-Mass Elements; IGE: Iron-Group Elements.} 
\end{figure}

For simulations C$_{8.91}$, O$_{9.46}$ and C$_{9.77}$ we include figures showing, as a 
function of the explosion time, the released nuclear energy, gravitational energy and 
kinetic energy (Fig.~\ref{fig:S6}) and the chemical composition (Fig.~\ref{fig:S7}). %
In these figures, we observe that the simulation with the highest initial
density, C$_{9.77}$, releases the most energy and burns the most material into
iron-group elements (IGE). It is also evident, however, that the simulation with
an off-centre ignition, O$_{9.46}$, burns more quickly initially and produces
more IGE during the first 1~second of the explosion compared to the other
simulations, consistent with the observed enhanced IGE mass fraction in the
remnant for this particular simulation (bottom panel of Fig.~4 in the main text). %
Further studies will be required to fully clarify the cause(s) of the faster
initial burning, which may reflect an earlier onset of turbulence, differences
in flame surface area, slight differences in temperature and density or a
combination of these. %

\begin{figure}[p]
\centering
    \includegraphics[width=\linewidth]{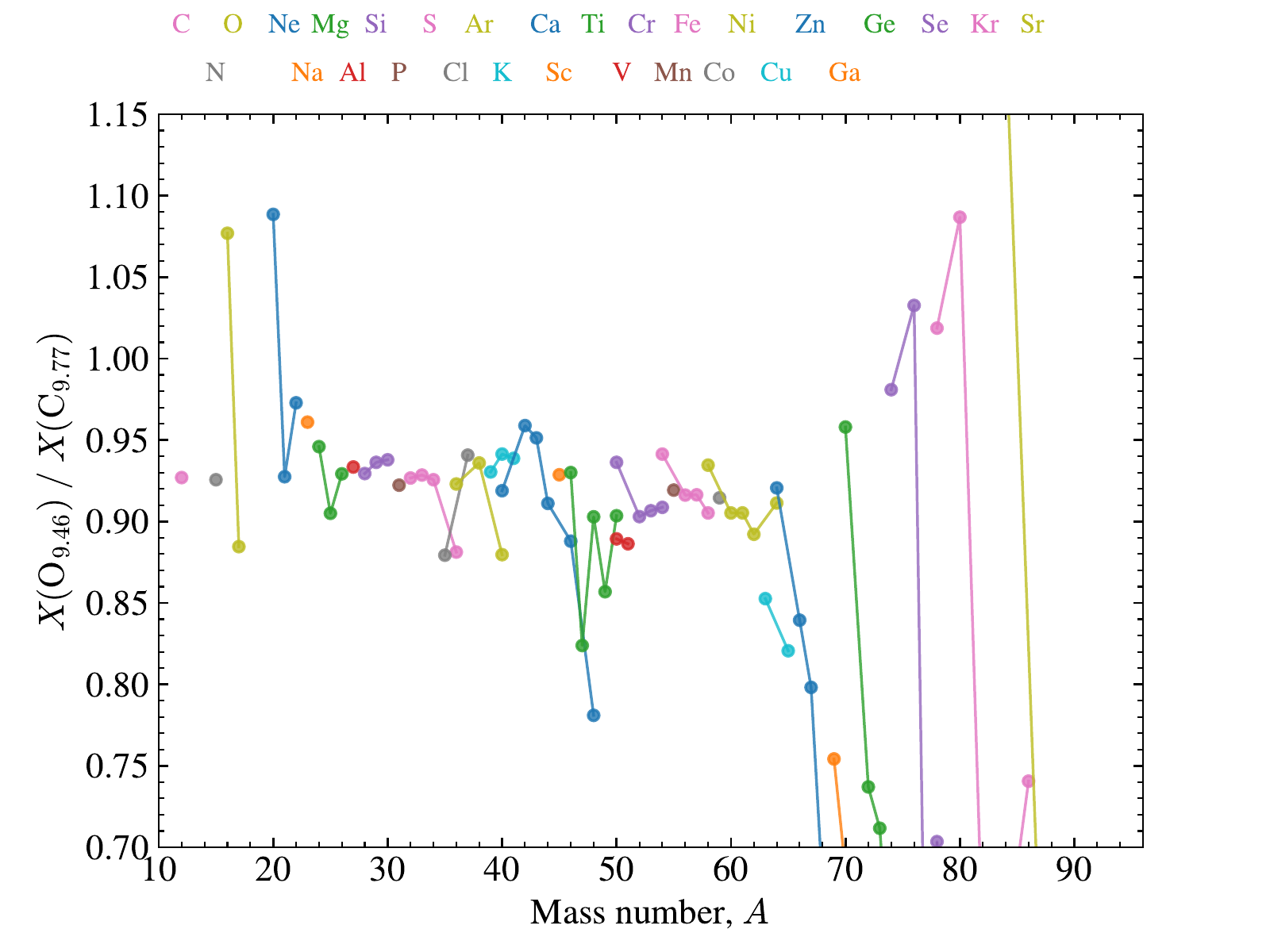}
    \caption{\label{fig:S8}
      Comparison of mass fractions of stable isotopes in the ejecta of
      simulations O$_{9.46}$ and C$_{9.77}$ after decaying for $0.32\times
      10^9$~yr. Only isotopes with mass fractions relative to solar of
      $X/X_{\odot} > 10^{-5}$ have been included.}
\end{figure}

Finally, we include a figure showing the mass fractions of stable isotopes in
the ejecta of O$_{9.46}$ relative to C$_{9.77}$ after decaying for $0.32\times
10^9$~yr (Fig.~\ref{fig:S8}), which illustrates the small, but significant
differences caused by the change in ignition density and geometry.

\input{Sams_table}

%% file: Sams_table.tex
 \begin{table}[p]
 	\centering
      \caption{Results from 3D hydrodynamic simulations of
        electron-capture ignited deflagration waves in dense ONe
        cores. The central density at ignition, ignition type,
        bound remnant mass, ejecta mass, remnant mass fractions of
        O+Ne and Fe-group elements, minimum electron fraction
        achieved in the simulation, average electron fraction in
        the bound remnant and effective Chandrasekhar mass of the
        bound remnant, are given.}
 		\label{tbl:hydro}
 		\begin{tabular}{c c c c c c}
 			\toprule
 			model &
 			resolution &
 			$\rho_\mathrm{c}^\mathrm{ign}$ &
 			ignition &
 			$M_\mathrm{rem}$ &
 			$M_\mathrm{ej}$ \\
 			& & ($10^9$ g cm$^{-3}$) & & ($M_\odot$) & ($M_\odot$) \\
 			\midrule
 			C$_{7.94}$ & $512^3$ & 7.943 & central & 0.438 & 0.951 \\
 			C$_{8.91}$ & $576^3$ & 8.913 & central & 0.366 & 1.027 \\
 			O$_{9.46}$ & $576^3$ & 9.462 & off-center & 0.291 & 1.103 \\
 			C$_{9.77}$ & $576^3$ & 9.772 & central & 0.245 & 1.150 \\
 			\bottomrule
 			\toprule
 			model &
 			$X_\mathrm{rem}^\mathrm{O+Ne}$ &
 			$X_\mathrm{rem}^\mathrm{``Fe"}$ &
 			$Y_\mathrm{e}^\mathrm{min}$ &
 			$\langle Y_\mathrm{e,rem}\rangle$ &
 			$M_\mathrm{Ch}^\mathrm{eff}$ \\
 			& & & & & ($M_\odot$) \\
 			\midrule
 			C$_{7.94}$ & 0.70 & 0.28 & 0.396 & 0.491 & 1.377 \\
 			C$_{8.91}$ & 0.68 & 0.29 & 0.373 & 0.489 & 1.366 \\
 			O$_{9.46}$ & 0.62 & 0.36 & 0.382 & 0.486 & 1.347 \\
 			C$_{9.77}$ & 0.66 & 0.32 & 0.350 & 0.485 & 1.345 \\
 			\bottomrule
 		\end{tabular}
 \end{table}